# ADS-Directory Services for Mobile Ad-Hoc Networks Based on an Information Market Model

Christian Hutter, Matthias R. Brust, and Steffen Rothkugel

University of Luxembourg[**]
Department of Sciences, Technology and Communication
L-1359 Luxembourg, Luxembourg
{Christian.Hutter, Matthias.Brust, Steffen.Rothkugel}@univ.lu

**Abstract.** Ubiquitous computing based on small mobile devices using wireless communication links is becoming very attractive. The computational power and storage capacities provided, allow the execution of sophisticated applications. Due to the fact that sharing of information is a central problem for distributed applications, the development of self organizing middleware services providing high level interfaces for information managing is essential. ADS is a directory services for mobile ad hoc networks dealing with local and nearby information as well as providing access to distant information. The approach discussed throughout this paper is based upon the concept of information markets.

## 1 Introduction

Today there are a large number of mobile devices like phones, personal digital assistants and tablet PCs available. Most of them are equipped with wireless communication adapters for local area communications. Storage capabilities and computational power are continuously increasing, allowing users to run even sophisticated applications, not only from the traditional domains like personal information management or business applications but also offering entertainment, fun and recreation.

In this regard, using ad-hoc networks as underlying topology is promising due to fostering ubiquity. Mobile ad-hoc networks are formed by large numbers of portable devices which might not have a fixed position and might join or leave the network at any time. Low bandwidth, unreliability of wireless links, small transmission ranges, and unpredictable network topology changes pose technically challenging problems. In realistic ad-hoc networks, resources and services need to be spread across a bigger area than just the transmission range of a single device. Hence, devices are required to communicate with one another by routing information through intermediate nodes, resulting in so-called multi-hop ad-hoc networks. Aside from that, nodes should also be considered as information providers. Users are supposed to collaborate with others, introducing the need to share and disseminate information across multiple nodes.

[**] This research is supported by the Luxembourg Ministre de la Culture, de l'Enseignement Suprieur et de la Recherche. Any opinions, findings and conclusions or recommendations expressed in this paper are those of the authors and do not necessarily reflect the views of the Luxembourg Ministre de la Culture, de l'Enseignement Suprieur et de la Recherche

This paper introduces ADS, a generic middleware Ad-hoc Directory Service, aimed at giving applications access to information located on more distant remote devices. The subsequent sections are organized as follows. Section 2 discusses several distinct ad-hoc application scenarios, in particular auction systems, ubiquitous multiplayer gaming, and M-Learning, motivating the usefulness of ADS. In section 3 a detailed description of main ADS components is given. Section 4 compares ADS to similar middleware solutions for mobile ad-hoc networks. Finally, section 5 gives a summary and provides ideas for future work.

## 2    Application Scenarios

In order to illustrate the usefulness of ADS, we present several application scenarios in mobile ad-hoc networks with different characteristics [1]. Firstly, we describe UbiBay, an auction system that can be considered as a counterpart to eBay for mobile ad-hoc networks. Secondly, we introduce a learning scenario, an ad-hoc forum, giving an example of an educational application for mobile ad-hoc networks. Finally, gaming applications are examined.

*Auction System.* UbiBay is an adaptation of prominent auction systems like eBay into a mobile ad hoc environment where users want to either sell items or bid on them. Auctions are run by agents that populate marketplaces. A marketplace is a geographical region where information is traded at given times. Users are not constrained to be at the marketplace physically, but are allowed to utilize other ones mobile devices located at the marketplace to let a software agent or a service installed on each device negotiate with others on their behalf [2]. Applications query those marketplaces to get a list of current auctions, allowing users to place bids.

*M-Learning.*  In an M-Learning environment students are working collaboratively to coordinate their effort. For instance before an exam, students should spend as much time as possible for their preparations. Thus they want to use the time between two lectures for studying. Unfortunately, students' agendas typically are very individual, making it extremely difficult for them to form common study groups. To still be able to join forces they might use an Ad-hoc Forum [3]. Students can use the according PDA driven application to submit their questions to the forum. While passing or meeting friends and fellow students, information like questions and answers are exchanged by the PDAs. The system aims at finding solutions for the user's questions or knowledge lacks. Users are able to evaluate replies to the questions. The results of this procedure are used to exclude non qualified answers.

*Gaming.*  While traveling together, people can use their mobile devices for multi-player gaming. For example a PDA might determine whether somebody nearby is interested in joining a game. As soon someone is found, the game integrates the new device into the gaming community. All participating PDAs thereby form an ad-hoc network. This scenario focuses mostly on groups of users located comparably close to each other like in a train wagon or bus.

## 3 ADS: An Ad-Hoc Directory Service

ADS is an approach for a fully distributed directory service for mobile ad hoc networks, based upon an information market model. It allows the sharing of potentially replicated information among applications running on the nodes of a mobile ad-hoc network. ADS is supposed to be part of a middleware available on each participating device. Applications are supposed to be used in a collaborative way. The data those applications operate on are initially generated on a particular device. This information might be useful for the local device only, or might be shared with other instances of the application-respectively with other users. In the latter case, the need for disseminating the information in a controlled way arises.

### 3.1 Information Markets

Due to the very dynamic structure of the network and the potentially high number of interacting devices, it is neither sensible nor possible to directly query any device. Furthermore, in ad-hoc networks there is no notion of central servers. Hence, strategies for collecting, exchanging, and gathering information are required. The approach proposed in this paper for tackling this problem is to exploit existing characteristics of real-life ad-hoc network infrastructure. In particular, it is possible and sensible to identify hot spots, where the density of devices can be expected to be above a certain threshold. Possible locations of such hot spots are central areas of cities, restaurants, cafeterias or entrances of buildings like train stations, warehouses or universities. Because environmental conditions might change during a day, however, locations of hot spots are not stable. Thus the ADS needs to be context-aware, adapting itself to the current situation. The student cafeteria of a university for instance might be well visited during daytime but deserted during nighttime or weekends. Based upon those hot spots, we encourage their use for what we call information markets, concentrating and managing large amounts of information in an adequate limited region and make it accessible to interested applications. Initially, the ADS knows the location of at least one information market. During runtime the platform discovers other markets, particularly their location and available information. Either the moving devices themselves disseminate this information in the network or it is propagated while communicating with an already known information market.

In order to make data available to applications, ADS supports two different types of queries: synchronous local and smart remote queries. Both types of queries are useful by themselves, albeit a combination of them might be sensible to be applied as well. Synchronous local queries are used to collect information from local and nearby devices. They are called synchronous because they follow a synchronous communication model. Asynchronous smart remote queries in turn enable applications to retrieve data from information markets. They are asynchronous in nature, because their execution time respectively behavior cannot be determined in advance.

### 3.2 Synchronous Local Queries

The main purpose of synchronous local queries is to collect information that is available immediately. Firstly, the information maintained on the local device by the ADS information broker component is retrieved. Optionally, the data might be augmented by query results from nearby devices. This is realized by sending queries to the information brokers in the immediate neighborhood. This information gathering process is based upon a timeout mechanism to cope with the properties of ad-hoc networks properly.

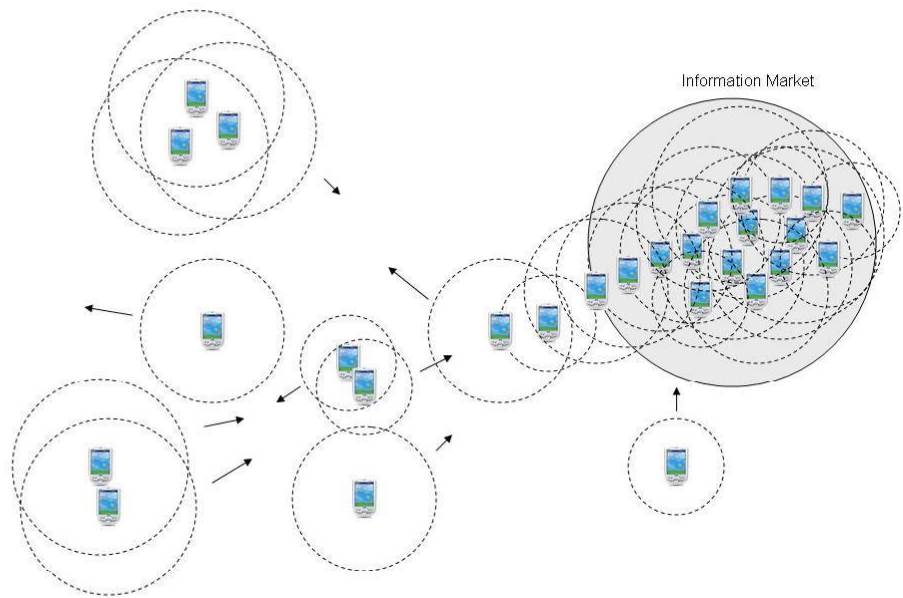

**Fig. 1.** Network scenario showing an example mobile ad hoc network including an information market

Taking the example of gaming, synchronous local queries provide a suitable abstraction. Interaction is restricted to the neighborhood. ADS might be used for instance to discover currently active games together with information about players involved. Additionally, slowly evolving game information e.g. related to trading can be handled appropriately by ADS synchronous local queries.

In the M-Learning scenario, this type of query is well-suited to collect information for instance while staying in the cafeteria. This way, knowledge from fellow students also having a break can be retrieved directly. Nevertheless, in that case it is also sensible to gather more information from other sources by referring to information markets. This is realized by the second type of queries: asynchronous smart remote queries.

### 3.3 Asynchronous Smart Remote Queries

Aside from retrieving information from the immediate neighborhood, it shall be possible to consult information markets as well. Queries launched will be sent to the information market, starting to collect results there. Queries might stay active for a given time, sending back results to the initiator in chunks. When transferring the query ADS must use smart forwarding strategies, so that the query finally will arrive at the information market. By tracking user movement it is possible to estimate the direction a device is heading to in the near future.

Queries as well as responses can contain meta-data. Mandatory meta-data of queries include information about the query initiator together with some contextual information about his planned movements—e.g. taken from his calendar—that is used to determine time and location of where to send results to. Optional meta-data of queries might for instance indicate an expected number of results. Knowledge about other information markets can be propagated and collected by including it in the meta-data of responses.

Taking the example of UbiBay, asynchronous smart remote queries can be used for example to gather information about ongoing auctions respectively items offered. In the Ad-hoc Forum scenario, questions as well as answers are collected and pooled. Hence, the number as well as the quality of the knowledge available at the market can be expected to be high, especially compared to the cafeteria case aforementioned.

### 3.4 Information Market Management

Information markets are supposed to manage potentially large amounts of data. This is done in a distributed and replicated way. The management strategies are described in the subsequent sections.

**The IMM.** The devices composing an information market need to act in a self-organizing way. When a device enters the information market, it contacts the information market manager (IMM). If there is no response, a new IMM is created on that device. The IMM collects the information about the free capacity available on the information market and about replicated information on each device. In case multiple IMMs are created at the same time, e.g. when several devices arrive simultaneously at an empty market, they need to select a leader using standard algorithms. The IMM assigns new information which arrives at the market to devices with free capacities. Depending on the level of importance of the information it might become replicated on several hosts. Devices to which the IMM already assigned data are called active while devices not holding data are passive to the information market.

The IMM needs to use load balancing algorithms to distribute information across the devices available, depending on their free capacities. As all information is added to the system by the IMM it can keep track about the type of information which is hosted on its market. This meta-information can be added to asynchronous smart remote queries to improve the chances of retrieving relevant information fast.

**Dealing with Mobility.** When a passive device leaves the area of an information market there is no loss of information. In case an active device *A* moves out of an information market its data needs to be transferred back to the market. *A* transfers the data to another device *B* which is still at the market or at least heading into the right direction. If *B* is already within the area of the information market it forwards the data to the IMM for redistribution. Otherwise it keeps the data until reaching the information market to finally forward it. If the direction of *B* changes before it enters the information market, B transfers the data to a different device heading towards the market.

A special case occurs if the IMM leaves the information market. However, the same strategy used by active devices when leaving the market is applied. Due to race conditions, a new IMM might have been created in between. In that case, the old IMM will deactivate itself. To increase the probability of the IMM staying inside the market it tries to reside in its center. If needed, the role of the IMM is taken over by a different device.

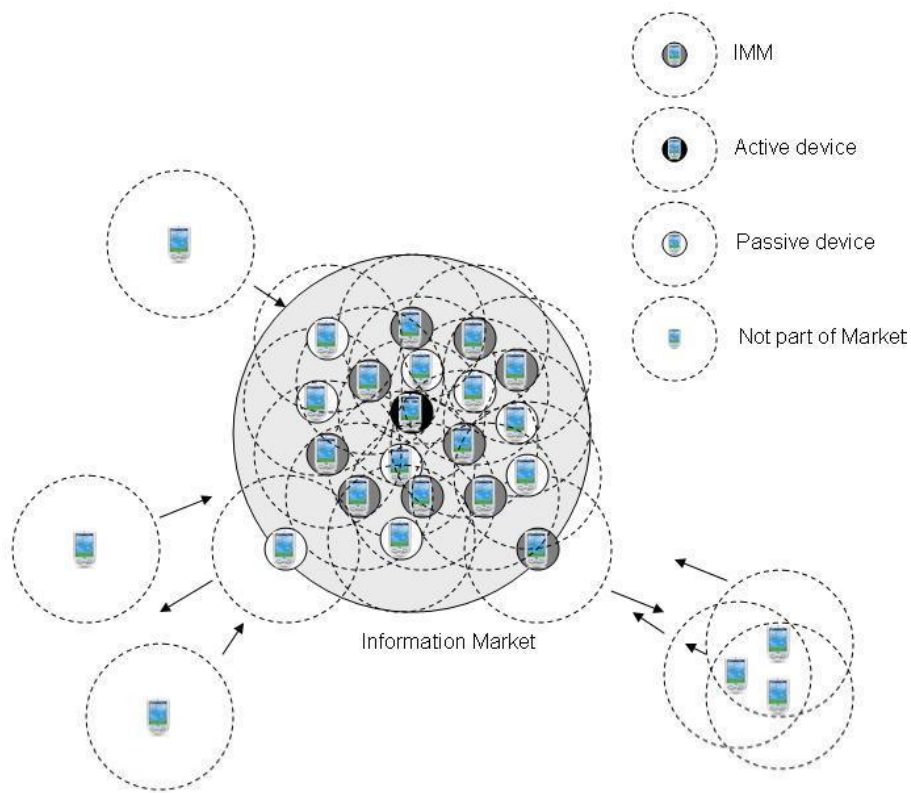

**Fig. 2.** Zoom to the information market showing (moving) active and passive devices and an IMM

**Communication within an Information Market.** Information markets differ from the rest of the ad-hoc network. Due to the high amount of available devices within a limited geographic area it is possible to establish a fast communication and information distribution between the participating devices. Even if they are not in direct communication range it is possible to use other nodes to forward messages using flooding techniques in this limited area. Messages sent to the IMM only need to be forwarded to the center of the market as the IMM is supposed to stay in this area.

**Disaster Recovery for Active/Passive Market Devices.** Disaster recovery needs to be done when a devices leaves a market without being able to perform the normal "leave market" actions mentioned above. This can happen as a result of a power-down or hardware respectively software error. In the case the device was passive to the market no information is lost. In the case of an active device, all information that is not replicated is also lost. The IMM can always try to restore replicated information of devices which are able of sending a final sign off message before shutting down. If a device can not send this message it will depend on the strategy chosen if the degree of replication can be reestablished (cf. also the paragraph describing information replication below).

**Disaster Recovery for the IMM.** The loss of the IMM will be detected when information will be added to the system or a new device enters the information market. At that time, a new instance of the IMM will be created. During initialization of the new IMM it asks all active or passive market devices about their free capacities. In case several IMMs are created, leader election techniques are used to elect one of them (c.f. the paragraph 3.4 about the IMM ). Basically, in the current design the IMM is a central entity. However, in case of failures, it is possible to create a new IMM dynamically. The state of the IMM can be reconstructed by gathering according data from the devices on the market.

**Publishing Information on an Information Market.** When an application or service wants to publish information, it needs to send it to an information market. The selection of a target market can be done context-aware by several strategies, e.g. by aspects of best fitting in terms of types of information already managed by a market or by distance to targeted applications. In both cases the information is forwarded to a market using the same strategies as applied in the asynchronous smart remote queries. The IMM handles the dissemination and replication of newly available data, taking into account meta-data like lifetime or replication degree.

**Information Replication.** The ADS allows information replication on information markets. If the IMM gets information of high importance, as indicated by meta-data determining the replication degree, it will assign this information to multiple devices. For reasons of recoverability those nodes need to know that they are now managing replicated information. To prevent information loss the IMM also keeps a list of the replicated information. The ADS provides two strategies to keep the number of copies at the suggested value, differing in the bandwidth usage. The first strategy requires devices to be able to send a short sign-off message when leaving a market. When the IMM receives such a message from devices hosting replicated information it starts a recovery operation, i.e. it gets and assigns a new copy of the information to another device. If a device is no longer available and was not able to send a sign-off message one copy will be lost. To prevent this, in the second strategy the IMM periodically checks if all devices involved into a replication process are still available. If necessary, the aforementioned recovery operation is triggered.

## 4 Related Work

There are a number of well known directory services for traditional networks environments like OpenLDAP and Microsoft's Active Directory. But due to the very dynamic structure of mobile ad hoc environments together with the restrictions in terms of processing power and storage capabilities, those are not suitable directly. ADS is explicitly designed for use in such environments.

The application scenarios introduced require information to be stored for further usage. A wide variety of different application-dependent information types needs to be managed. Simple strategies like flooding the network by disseminating data using adaptive protocols like introduced in [4] are not reasonable in general. A different approach is used in NOM [5] which forwards queries and creates responses from every node in the network. In both strategies, broadcast storms are likely to occur. SIDE Surfer [6] in turn only allows the automatic exchange between directly connected devices based on user profiles, giving the applications access to a limited set of information only. ADS allows applications to use the synchronous local queries to access such information and additionally provides asynchronous smart remote queries for retrieving data from information pools currently available on information markets.

Both TOTA ("Tuples on the air" [7]) which aims at supporting adaptive context-aware activities and MeshMdl [8] use the tuple paradigm in mobile ad hoc networks. But they miss the concept of information replication which increases the probability of loosing important information.

## 5 Conclusion and Future Work

ADS is a middleware service designed for use in mobile multi-hop ad-hoc environments. It enables applications to directly share information with as well as receive information from the devices in the neighborhood, together with providing access to information markets. The main idea of introducing information markets is to be able to identify well-known places where different kinds of information from multiple applications can be pooled and exchanged. Within information markets, sophisticated algorithms can be applied e.g. in terms of load balancing and fault tolerance by replication. The interaction with information markets is facilitated by so-called asynchronous smart remote queries, which are long-term queries. After being launched, these queries travel to information markets, gathering information there to be sent back to their initiator. This makes the information market model well-suited for data that remains stable for a certain time.

The environment envisaged is neither stable nor fully predictable. The system might for instance suffer from network partitioning, resulting e.g. in queries reaching information markets either late or possibly not at all. Another problem occurs if the number of participating devices drops under a certain threshold. Then, proper dissemination and replication of information within information markets cannot be guaranteed anymore.

Strong guarantees cannot be given in such kind of environments anyway. It is only possible to minimize the impact of different kinds of failures and shortcomings. In the model proposed, this is done through the use of information markets together with their management strategies. As shown in [2], the information market model is a promising approach.

In the future, several interesting aspects need to be evaluated. For instance, the management of information about marketplaces themselves needs to be tackled, like their establishment and discovery. One possible solution here is to rely on general knowledge which locations might be well-suited as e.g. train stations and shopping malls. A different approach is to dynamically determine marketplaces by observing devices and discovering common patterns. The management of data within information markets is another area for future research, including for example to de-centralize the Information Market Manager. It might also be interesting to evaluate if an exchange of information across several information markets would improve the overall system effectiveness. Additionally, we currently assume information to be stored atomically. In order to leverage both fault tolerance and load balancing, it could be interesting to split one piece of information into multiple chunks obeying a certain level of redundancy, and to distribute those on several devices. Finally, several application domains exist that obey different interaction paradigms. Examples include cases where data becomes out-of-date rather quickly, or applications that are comparatively tightly coupled. Hence, other concepts respectively communication patterns aside from the information market model for sharing and exchanging information need to be developed.